\documentclass[graybox]{article}

\usepackage{graphicx}

\usepackage{todonotes}

\usepackage{natbib}
\bibpunct{(}{)}{;}{a}{}{,}  % adjust author-year citation format 

\usepackage[stable]{footmisc}

\usepackage{hyperref}
\hypersetup{colorlinks,
            linktocpage,%  make page number, not text, be link on TOC, LOF & LOT
            breaklinks,%
            citecolor=blue,%
            filecolor=blue,%
            linkcolor=blue,%
            urlcolor=blue,}

\usepackage[font=small,labelfont=bf]{caption}

\usepackage[tight,nooneline,FIGTOPCAP,bf]{subfigure}
\renewcommand{\thesubfigure}{\alph{subfigure}}
\makeatletter
    \renewcommand{\@thesubfigure}{(\thesubfigure)}  % subfigure labels:     (a)
    \renewcommand{\@@thesubfigure}{\thesubfigure}   % produced by \subref:   a
    \renewcommand{\p@subfigure}{\thefigure}         % produced by \ref:   3.1a
\makeatother

\usepackage{fancyhdr}
\usepackage{authblk}
\usepackage{combelow}
\usepackage{algorithm}
\usepackage{algpseudocode}
\usepackage{enumerate}
\usepackage{amssymb,amsmath}
\usepackage{moreverb}
\usepackage{listings}
\usepackage{multirow}
\usepackage{verbatim}
\usepackage{epsfig}
\usepackage{epstopdf}
\usepackage{makecell}

\usepackage[T1]{fontenc}% http://ctan.org/pkg/fontenc

\renewcommand{\thefigure}{\thechapter \arabic{figure}}
\makeindex             % used for the subject index

\begin{document}

% \title{Sample contributed chapter}
% % Use \titlerunning{Short Title} for an abbreviated version of
% % your contribution title if the original one is too long
% \author{Michael Gowanlock$^{\rm a}$\vspace{6pt}, and Ian S. Morrison$^{\rm b}$\\
% $^{a}${\em{Massachusetts Institute of Technology, Haystack Observatory, Westford, MA, U.S.A. \\E-mail: gowanloc@mit.edu}
% }\\
% $^{b}${\em{Affiliation author 2, E-mail: author2@email.com}
% }
% }

\title{The Habitability of our Evolving Galaxy}
% Use \titlerunning{Short Title} for an abbreviated version of
% your contribution title if the original one is too long
\author{Michael G. Gowanlock$^{\rm a,b}$$^{\ast}$\thanks{$^\ast$Corresponding
author.\vspace{6pt}}~~and Ian S. Morrison$^{\rm c}$\\
$^{a}${\em{Massachusetts Institute of Technology, Haystack Observatory, Westford, MA, U.S.A.}}\\
$^{b}${\em{School of Informatics, Computing, and Cyber Systems, Northern Arizona University, Flagstaff, AZ, U.S.A. E-mail: Michael.Gowanlock@nau.edu}}\\
$^{c}${\em{Centre for Astrophysics and Supercomputing, Swinburne University of Technology Hawthorn, Australia. E-mail: imorrison@swin.edu.au}}}

\date{}
\maketitle

\begin{abstract}
The notion of a Galactic Habitable Zone (GHZ), or regions of the Milky Way galaxy that preferentially maintain the conditions to sustain complex life, has recently gained attention due to the detection of numerous exoplanets and advances made in understanding habitability on the Earth and other environments.  We discuss what a habitable environment means on large spatial and temporal scales, which necessarily requires an approximated definition of habitability to make an assessment of the astrophysical conditions that may sustain complex life.  We discuss a few key exoplanet findings that directly relate to estimating the distribution of Earth-size planets in the Milky Way.  With a broad notion of habitability defined and major observable properties of the Milky Way described, we compare selected literature on the GHZ and postulate why the models yield differing predictions of the most habitable regions at the present day, which include: (1) the majority of the galactic disk; (2) an annular ring between $7-9$ kpc, and (3) the galactic outskirts.  We briefly discuss the habitability of other galaxies as influenced by these studies.  We note that the dangers to biospheres in the Galaxy taken into account in these studies may be incomplete and we discuss the possible role of Gamma-Ray Bursts and other dangers to life in the Milky Way.  We speculate how changing astrophysical properties may affect the GHZ over time, including before the Earth formed, and describe how new observations and other related research may fit into the bigger picture of the habitability of the Galaxy.

\end{abstract}

\section*{Keywords}
Astrobiology, Galactic Chemical Evolution, Galactic Habitable Zone, Habitability, Milky Way Galaxy.

\section{Introduction}
Astrobiology has developed significantly over the past decades through the advances made in multiple fields.   The search for and detection of extrasolar planets has made possible estimates of the number of planets in the Milky Way \citep{2010Sci...327..977B,Petigura2013}. New observational constraints have permitted modeling efforts to constrain the conditions necessary for planets to host a biosphere~\citep{2016A&A...592A..36G}.   The Solar System has been used as an analog to understand the possible conditions for life on other planets.  And beyond planets, other bodies such as moons \citep{1997williams,2010kaltenegger,2012quarles} have been suggested as potentially able to host a biosphere.   Furthermore, the life sciences have made inroads into understanding the origin and constraints for life on Earth \citep{2008martin,2006ehrenfreund}.  Combining the findings in these fields and others, new assessments of habitability on larger spatial and temporal scales are now possible. 

The Earth hosts many habitable environments, but the current conditions are possibly due to the presence of liquid water on the planet's surface.  The circumstellar habitable zone (HZ)~\citep{1993kasting,2013kopparapu} constrains the distance from a host star such that water remains in a liquid state on the surface of a planet having sufficient atmospheric pressure.  Indeed, our Solar System is a prototype for an exoplanetary system, where the Earth is found to be roughly within the middle of the HZ. Beyond the Solar System,  exoplanet detections suggest that terrestrial planets in the HZ are common within the Milky Way~\citep{Petigura2013}.

Although the study of habitability beyond the Earth is still in its early stages, there are astrophysical constraints on the suitability of planets to have the carrying capacity to host life over time within the Milky Way.  Studies have examined various aspects of the habitability of the Milky Way~\citep{2001Icar..152..185G,2004Sci...303...59L,2008SSRv..135..313P,2011AsBio..11..855G,2014MNRAS.440.2588S,2015ApJ...810L...2D,2015AsBio..15..683M,2015arXiv151101786F,2016ApJ...826L...3T,2016MNRAS.459.3512V,2016A&A...592A..96G} as increasing consensus on astronomical observational constraints have been realized over the past few decades.  Some of these works define the habitability of the Milky Way by spatial locations suitable for life over time, thus giving rise to the term the Galactic Habitable Zone (GHZ).  Some of the aforementioned references will be reviewed within this chapter.

Largely due to the many uncertainties involved in assessing the habitability of the Milky Way, the astrophysical processes that may lead to or are detrimental to life necessitate a coarse-grained view of the factors that make the Milky Way a habitable environment.  Furthermore, the discussion of habitability often assumes an implicit definition of what constitutes an environment suitable for life.  In terms of the GHZ, a habitable environment is considered one that is not frequently exposed to transient radiation events, where these events can deplete ozone in planetary atmospheres, exposing life to harmful radiation from its host star. However, this implies a habitable environment for surface dwelling life that would be particularly exposed to radiation from its host star.  Thus, existing life in other potentially protected environments (e.g., the ocean floor) are exempt from any analysis of the effects on their environments.  Furthermore, the study of the GHZ makes no claims on where preexisting life may be within the Galaxy; rather it is an assessment of the suitability for land-based life given the astrophysical mechanisms known to limit or bolster what are believed to be habitable environments. Given the uncertainties involved in this field of study, the habitability of the Milky Way is still largely an open question.

In this chapter, we review the concept of habitability on large spatial and temporal scales, we illuminate some of the key advancements made by the exoplanet community, we give an overview of the differing predictions and models of the habitability of the Milky Way, discuss the habitability of other Galaxies, revisit types of transient radiation events and their implications for the Milky Way, extrapolate and speculate on the habitability of the Galaxy before the formation of the Earth and finally conclude the chapter and discuss future research directions in the emergent field of habitability on large spatiotemporal scales.

\section{Habitability}\label{sec:habitability}
Before discussing the details of the habitability of our evolving galaxy, the Milky Way, we will discuss what is meant by a habitable environment.  The term habitability is widely used in differing contexts across the constituent research communities that make up the field of astrobiology.  The study of the habitability of the Earth starts with determining a list of the items that are common to all life on our planet.  There are broadly three major requirements for life as we know it: 1) an energy source~\citep{McKay2004}, such as chemical energy or the Sun (for life that requires photosynthesis); 2) a range of elements, C, H, N, O, P, S~\citep{Falkowski2006}, as all life on Earth is composed of these elements as they make up biomolecules; and 3) water to facilitate chemical interactions~\citep{Mottl2007}.  These appear to be necessary requirements for all life on Earth and are an accepted part of the definition of habitable environments on Earth.  However, life on other bodies may be able to thrive without these specific ingredients. For instance, it is possible that a solvent other than water may be used to facilitate chemical interactions.  Despite the many uncertainties inherent in applying our knowledge of the habitability of the Earth to other environments, having a minimal list of requirements for life on Earth forms the basis for understanding the conditions for life in other environments.

If we take a closer look at some of these requirements, it becomes clear that there are many prerequisites for these conditions.  For instance, having a solvent such as water implies that the temperature on a planet must be within a range such that water does not freeze or evaporate.  This means that the planet must have a suitable average orbital distance from its host star, and an atmosphere that can provide those conditions.  The atmospheres of Venus and Mars are two cases that cannot provide liquid water on their respective surfaces, where the Venusian atmosphere is too dense and traps heat, and the Martian atmosphere is too depleted and cannot retain enough heat at the surface causing freezing, or the water rapidly evaporates.  Thus, there are many interesting questions regarding the combination of the orbital radius and atmospheric composition required for planets to retain liquid water on their surfaces~\citep{2013kopparapu}.  Exoplanets likely have similar requirements to be habitable, and to remain so over long timescales.

From the previous discussion, we have made the leap from a rudimentary list of requirements for life on Earth, to speculating about how these requirements may lead to defining particular \emph{habitable environments}.  This opens up many questions about the limits to habitability on Earth, the formation of planets and their configurations in planetary systems, and the galactic environments that may bolster or limit the carrying capacity for life.  Exoplanets and their configurations have been detected through radial velocity and planetary transit methods~\citep{1998saar,2000Charbonneau}. These observations are the outcome of planet formation which occurs in protoplanetary disks~\citep{1996pollack,1997boss,2007Matsuo}. Modeling the formation of planets from protoplanetary disks attempts to reconcile fundamental physical processes with those planetary configurations that are similar to the Solar System and exoplanetary systems.  In combination with observations, they have shown that the (previously thought) exotic exoplanetary systems must be common throughout the Milky Way, and furthermore, that there are a wide range of possible planetary configurations that are significantly different than the Solar System.

By shifting the focus from the requirements for life on Earth to those environments that can support life, we implicitly ignore that the origin of life must occur, and that life itself influences its environment. For instance, the first forms of life on Earth used an anaerobic instead of aerobic metabolism.  However, once oxygen was plentiful in the atmosphere as a by-product of photosynthesis, new aerobic forms of life began to develop and thrive.  Utilizing oxygen yields more energy from the metabolic process, and as a result, life is more efficient in its homeostasis. This suggests that biology is rooted in physical processes, and that life having a tendency to change its environment is not a random process.   The build up of oxygen in the Earth's atmosphere is accounted for by photosynthesizing cyanobacteria~\citep{2005catling}. Without an increase in atmospheric oxygen, it is possible that only anaerobic organisms would still be in existence, and the evolutionary paths of ancient organisms would have been significantly different. Therefore, it is very likely that animal life, particularly land-based animal life, would not exist today, had it not been for life affecting its environment.  This necessarily presents a problem for the study of habitability.  If habitable conditions are required for life, but life is required for maintaining long term habitable conditions, then how can we study habitability without first understanding the conditions for the origin of life? Other works discuss this problem~\citep{2005catling,lineweaver2012,chopra2016}, and we refer the reader to those references and other chapters in this book, such as that by Chopra \& Lineweaver, for additional discussion.  When it comes to the study of habitability, we must assume that we are interested in maintaining the conditions for life for a sufficient timescale. Furthermore, it is necessary to be specific about the type of life and requisite environment that is of interest.  For instance, the Galilean moon Europa that orbits Jupiter is a target of interest because it may contain liquid water under its frozen water ice crust. Its source of heat is thought to be tidal heating as a result of its orbit around its parent body.  Constraining habitable conditions on Europa equates to understanding subsurface life.  The study of habitability needs to be mindful of these constraints, the influence that life has on its environment, and the numerous uncertainties that result when applying our knowledge of life on Earth to other bodies in the Solar System or Galaxy. With that in mind, we now move on to the habitability of our evolving Galaxy.

We may think that there are many requirements for life on Earth because we can study the numerous environments in which life is found in the biosphere.  However, when assessing habitability at larger scales, many of these prerequisites cannot be quantified or observed. For instance, exoplanet searches may claim a habitable planet detection based on its location in the HZ, without knowledge of its atmospheric characteristics or other factors that may make the planet habitable.  Unfortunately, until future missions can make measurements that detect biosignatures or other factors that constrain habitability, we are limited to a coarse-grained view of habitability throughout the Milky Way.  The GHZ is an idea that is similar to the HZ, initially defined by an  annular region of the Milky Way that can host habitable planets that are not too close to the center of the Galaxy or the outskirts.  The inner edge has been thought to be constrained by a high frequency of transient radiation events, and the outer edge constrained by insufficient metallicity for planet formation~\citep{2001Icar..152..185G} and these factors will be discussed in greater detail in upcoming sections.  By these constraints, a habitable planet is one that is considered to maintain an environment free from transient radiation events for timescales commensurate with complex life, or more particularly land-based surface dwelling life.  Therefore, the definition of the GHZ specifically considers a particular environment for life (the surface). This is because it is not clear that a transient radiation event would have a catastrophic impact on life found in the subsurface or oceans.  In all that follows, when we discuss the habitability of a planet in the context of the Milky Way, we are referring to those conditions that may support land-based complex life.  The region of habitability, or ``galactic habitable zone'' refers to the regions in the Galaxy over time that have the highest carrying capacity for complex life over time.  Examining the effects of astrophysical events on the habitability of planets contributes a small part to our understanding of the habitability of the Milky Way throughout cosmic time, and our precarious situation as a species on our planet.

\section{The Exoplanet Era}
%section on better observational constraints

The first exoplanet was detected in 1992 orbiting a pulsar~\citep{1992Natur.355..145W}.  Since then, at the time of writing, there have been over 3,000 additional exoplanets confirmed\footnote{\url{http://www.exoplanet.eu}, 3610 planets as of May 17, 2017.}.  The \emph{Kepler} mission~\citep{2010Sci...327..977B} has detected the majority of these exoplanets through the photometric transit technique.  Combining transit and radial velocity measurements, the orbital period, planet radius and mass can be constrained, thereby providing insight into the configurations of other planetary systems. These detections have given rise to an intense period of research on exoplanets. Statistical constraints across the sample of detected planets have been conducted to estimate the fraction of HZ planets around Sun-like stars~\citep{Petigura2013}.  The mass, radius, and orbital period have been used as parameters into models of a planet's atmospheric composition to determine whether it can retain liquid water on its surface~\citep{2016AsBio..16..443S}.  Furthermore, the study of atmospheric biosignatures may yield evidence for the existence of life on a planet~\citep{2003AsBio...3..471P,2005AsBio...5..706S,Kaltenegger2007}.  These avenues of research have combined the efforts of astronomy, planetary science, geology, biology, and other fields, to determine the conditions for potential life on other planets.  Future missions and observing campaigns will provide much needed insight into whether some of the detected HZ exoplanets are in fact habitable, and will guide models that constrain the habitability of planets around their host stars.

The two techniques that have indirectly detected the majority of the exoplanets are the transit and radial velocity techniques. When a planet's orbit passes through the line of sight between the observer and its host star, the transit of the planet will cause a small drop in the observed brightness of the star that can be used to infer the presence of the planet.  This technique yields the planet's radius and orbital period.  With regards to the radial velocity technique, as a planet orbits its host star, the gravitational effect on the star can be measured. By examining Doppler variations in the stellar spectrum, the star will appear to be periodically moving towards and away from the Earth.  This method is sensitive to the relative masses of the planet and host star, favoring larger planets orbiting lower mass stars.  Combining findings from both the transit and radial velocity methods, a planet's orbital period, size, and upper limit on mass can be obtained.  With the mass estimate and size, the bulk densities can be calculated, and chemical compositions inferred.  From the study of large numbers of exoplanets, statistics can be gathered that can be directly applied to understanding the habitability of the Galaxy.  We now review some of these key findings.

We now know that the majority of stars are likely to host planets. \cite{Mayor2011} finds that 75\% of solar type stars host a detectable planet with an orbital period less than 10 years. This estimate does not extrapolate to those planets that are not detectable due to observational constraints; therefore, this may be a lower limit on the fraction of solar type stars with planets.

\begin{figure*}[ht]
\begin{center}
 \includegraphics[width=0.7\textwidth]{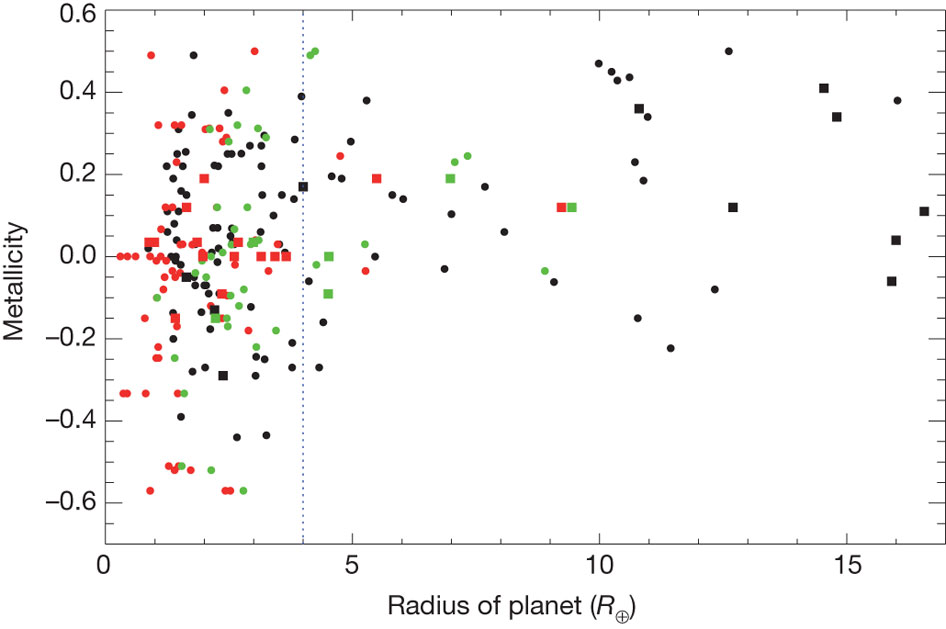}
  \caption{Figure 3 from \cite{2012Natur.486..375B} that plots host star metallicity vs. planet radius in Earth radii ($R_\oplus$). Small planets form around stars with a range of metallicity values, and large planets are strongly correlated with high metallicity. The mean metallicity for small size planets is near the solar value. The vertical dotted line separates the two planet sizes in the sample at $4.0R_\oplus$.  For more information on the sample of planets and the representation of figure markers that refer to single planet systems, and planet sizes in planetary systems with multiple planets, see~\cite{2012Natur.486..375B}.} 
  \label{Fig3_from_2012Natur.486..375B}
\end{center}
\end{figure*}

Properties of host stars are correlated with the type of planets that are likely to orbit them.  Before large scale transit searches, the radial velocity method was the primary means for detecting exoplanets, and this method has those aforementioned observation biases that mean it is able to more easily detect large planets.  \citet{2005ApJ...622.1102F} found a planet-metallicity correlation that shows gas giant planets are strongly correlated with host star metallicity, where metallicity refers to the abundance of elements in the star that are not hydrogen or helium. From the \emph{Kepler} mission, \citet{2012Natur.486..375B} find that the correlation between small planets and metallicity is flat, which means that small planets form from protoplanetary disks that have a range of chemical abundances, and do not require a significant abundance of metals for small planet formation.  Therefore, small Earth-sized planet formation is moderately independent of chemical evolution, thus Earth-sized planets will be common in the Galaxy. Figure~\ref{Fig3_from_2012Natur.486..375B} shows the host star metallicity as a function of planet radius from \citet{2012Natur.486..375B}. From the figure, it can be seen that small planets form around stars with a wide range of metallicities, and larger planets are correlated with higher metallicities, consistent with \citet{2005ApJ...622.1102F} and \citet{Mayor2011}.  

The observations of planet size and metallicity are consistent with the core accretion model of planet formation.  Gas giant formation first needs to produce a large rocky core in the protoplanetary disk that then has a runaway gas accretion period~\citep{1996Icar..124...62P}. This explains why gas giants are primarily found around metal-rich host stars. In comparison, smaller rocky planets do not need a large abundance of planetary building blocks (high metallicity) in the protoplanetary disk.  These findings have direct implications for the habitability of the Milky Way, as the metallicity of the Galaxy has evolved with time, and this evolution will influence how planets form over time.  

To constrain the habitability of planets, and not just their abundance around stars of various masses and chemical compositions, determining the fraction of Earth-size planets that orbit within the HZ of their host stars is required. \cite{Petigura2013} finds that 22\% of Sun-like stars have Earth-size planets orbiting in their HZ.  This suggests that, of the many billions of Earth-size planets predicted in the Milky Way, a significant fraction may have conditions favorable for life.

\section{The Habitability of the Milky Way} \label{Section_Hab_MW}

In this chapter, we have discussed habitable zones on the Earth, Solar System, and the Galaxy.  The Galactic Habitable Zone originated as an idea that was analogous to the HZ~\citep{2001Icar..152..185G}, and thus it was assumed to be annular in nature, with sharp inner and outer limits to habitability.  However, as will be examined in this section, an annular perspective may not necessarily be reasonable for the Milky Way. Therefore, when discussing the morphology of the GHZ, we do not mean for the reader to assume an annular region, such that we are not predisposed to a goldilocks view of habitability.

When considering the habitability of the Galaxy, a coarse-grained view of habitability is assumed, as astrophysical factors that may inhibit or advance life are considered.  Some astrophysical prerequisites for life include a star, with a rocky planet in the HZ, that can withstand transient radiation events, such as supernovae (SNe), gamma-ray bursts (GRBs), or active galactic nuclei (AGN) over a sufficient time period that allows for land-based complex life to evolve and survive.  As mentioned in the discussion regarding habitability, an implicit assumption must be made that the conditions that allow complex life to thrive must exist, even though those environments may have only been made possible through life transforming its environment.  However, when future observations are better able to constrain the fraction of truly habitable planets, those statistics will be able to be used as input into newer models of the habitability of the Milky Way.        

%overview on all of the works:

Through stellar nucleosynthesis, the generations of (mostly high mass) stars produce the materials needed for planet formation.  The initial generations of stars would have formed nearly entirely out of hydrogen (and helium) without any of the heavier elements to produce planets in protoplanetary disks. The metallicity of the Milky Way evolves with time; therefore, the abundance of planet-building material has increased over time in the Galaxy. \cite{2001Icar..152..185G} proposed quantifying the metallicity selection effect to constrain the GHZ. Using constraints on the chemical evolution of the Galaxy, and an estimate of the abundance of metals needed to form a habitable planet (half the Solar value), the habitablity of the Milky Way can be assessed. \cite{2001Icar..152..185G} report that metal-poor regions of the Milky Way, such as the halo, thick disk, and outer thin disk (at higher galactocentric radii), may not have enough metals for planet formation.  Furthermore, they compare the metallicity of the Milky Way to other galaxies. Since higher luminosity galaxies are likely to be more metal-rich (luminosity traces star formation, and star formation traces metal abundance), they find that the Milky Way, being more luminous than other galaxies in the local Universe, is likely to host more planets.  \cite{2001Icar..151..307L} uses the metallicity evolution and star formation rate of the Universe to quantify the age distribution of planets in the Universe over cosmic timescales.   \cite{2001Icar..151..307L} finds that the majority of Earth-like planets in the Universe are 1.8$\pm$0.9 Gyr older than the Earth.  Furthermore, he suggests that Hot Jupiters (gas giants orbiting near their host stars) would inhibit the formation of Earth-like planets, as they may accrete the material required for terrestrial planet formation. Since these Hot Jupiters are found around metal-rich host stars, then there may be a selection effect, where high metallicity inhibits rocky planet formation, and low metallicity inhibits planets from forming entirely.

\cite{2004Sci...303...59L} constrained the habitability of the Milky Way as a function of both metallicity and catastrophic transient radiation events. They find that the habitability of the Milky Way is described in terms of an annular region, where at the present day, the inner edge of the GHZ is defined by too high of a frequency of lethal SN events, and too high metallicity that produces Hot Jupiters \citep[as discussed above in ][]{2001Icar..151..307L}, and the outer edge is defined by insufficient metallicity for planet formation.  \cite{2004Sci...303...59L} finds that the GHZ is between 7--9 kpc in galactocentric radii that widens with time (the radius of the disk of the Milky Way is approximately 15 kpc). Additionally they find that 75\% of the stars in this region are older than the Sun. \cite{2001Icar..152..185G} and \cite{2004Sci...303...59L} set the stage for constraining habitability on larger spatial and temporal scales, and carefully considered these calculations largely before many exoplanets had been detected.

With the concept of the GHZ described as an extension to the HZ, \cite{2008SSRv..135..313P} notes that the question of habitability on galactic scales may be too difficult to quantify, because the prerequisites for habitability are not well established.  Despite this perspective, \cite{2008SSRv..135..313P} models the star formation history, rocky planet formation, presence of Hot Jupiters, and SN events, and finds that the entire disk of the Milky Way may be habitable.  Furthermore, they find that the inner Galaxy is likely more habitable than the outer Galaxy at later epochs. Early epochs would have had too many SNe, but after these lethal events become less frequent, there are likely to be more Earth-like planets in the region than at the solar neighborhood or outskirts.  Furthermore, our take-away from \cite{2008SSRv..135..313P} is that the framework of a GHZ may be insignificant because the entire disk may be habitable, the physical processes underlying habitability are not well constrained, and the results of such studies are largely dependent upon model assumptions.

\begin{figure}[htp]
\begin{center}
 \includegraphics[width=0.7\textwidth]{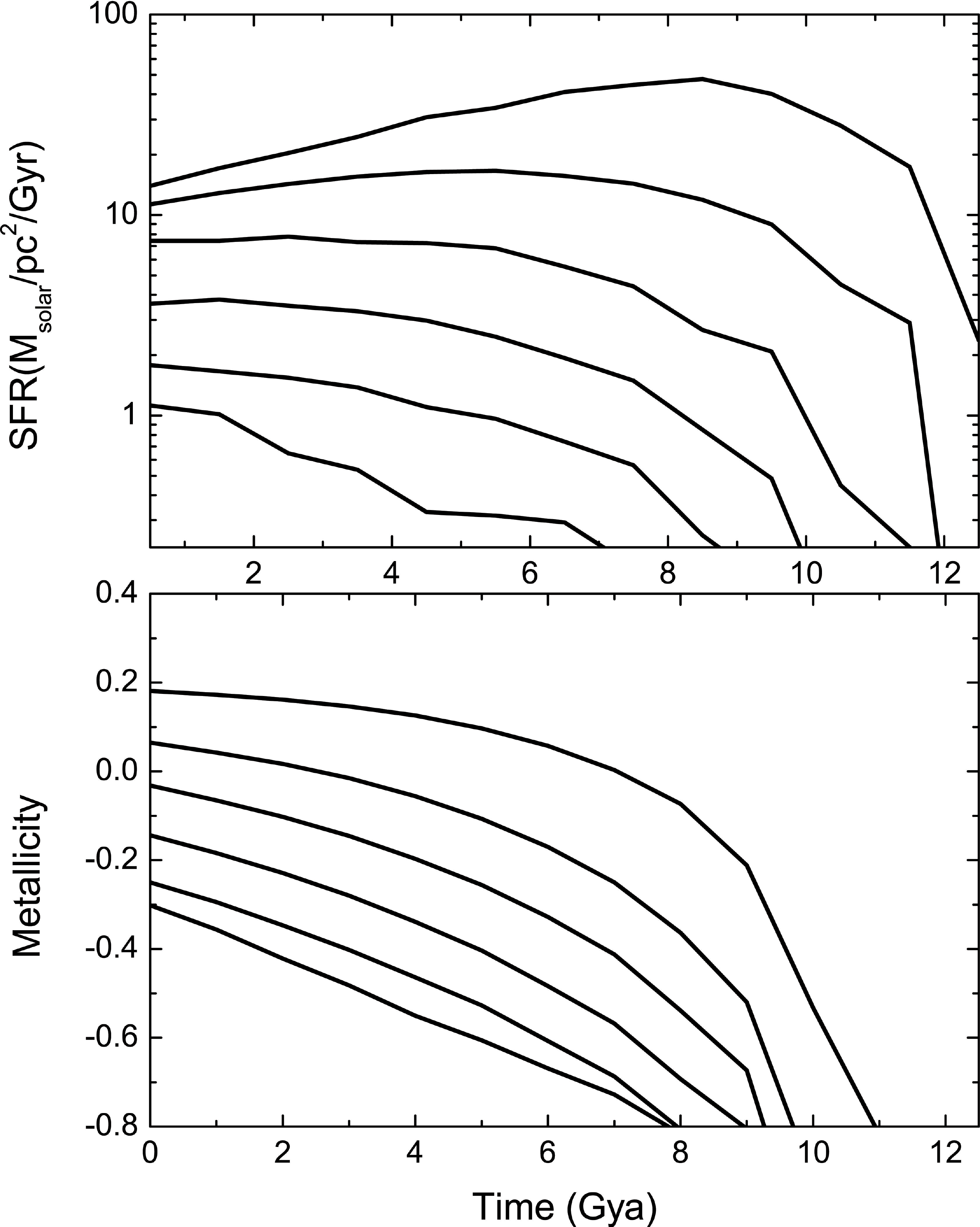}
  \caption{The star formation rate (upper) and metallicity (lower) as a function of time in Model~4 from~\cite{2011AsBio..11..855G}.  The curves represent the values at six galactocentric regions as ordered from top to bottom: 2.5, 5, 7.5, 10, 12.5, 15 kpc.  There is a burst of star formation  in the inner Galaxy that declines with time (upper). Due to the inside-out formation history, the inner Galaxy produces metals earlier than the outskirts (lower). Note that the present day is 0 Gya.} 
  \label{Fig1_from_2011AsBio..11..855G}
\end{center}
\end{figure}

Our previous work, \cite{2011AsBio..11..855G} used the same prerequisites for complex life as \cite{2004Sci...303...59L}.  We considered the formation of rocky planets in the HZ of their host stars that are free from SN events over a timescale sufficient for the rise and long-term stability of land-based complex life.  In particular, using a Monte Carlo approach, we populated stars using the 3-dimensional stellar number density of the thin and thick disks~\citep{2008ApJ...673..864J}, then for each star we applied the following properties: a mass from the initial mass function (IMF)~\citep{1955ApJ...121..161S,2001MNRAS.322..231K}, a stellar lifetime (from the IMF), a birth date from the inside-out star formation history, and a metallicity, as a function of radius and time~\citep{2006MNRAS.366..899N}.  With this model of the Milky Way, we then assigned planets to stars as a function of metallicity. Then, in this population of disk stars, we assigned them a probability of becoming a SN, either type II (SNII) or type Ia (SNIa).  SNII occur from massive stars; therefore, they occur primarily in star-forming regions and are young when they occur, and single degenerate SNIa are thought to occur in binary star systems that span a distribution of ages~\citep{2008ApJ...683L..25P}.  Therefore, star forming environments are likely to host SNII, and SNIa can occur independently of recent star formation.
Using the inside-out star formation history from \cite{2006MNRAS.366..899N}, stars initially form in the inner Galaxy and star formation propogates outwards over time.  Therefore, the average age of stars (and planets) in the inner Galaxy is greater than the outer Galaxy. Figure~\ref{Fig1_from_2011AsBio..11..855G} shows the star formation history and metallicity evolution of Model~4 in \cite{2011AsBio..11..855G}. From the figure, there is a burst of star formation that declines with time.  The inner Galaxy hosts more stars and has more chemical evolution than at the solar radius ($R_{\odot}=8$ kpc) or the outskirts.

\begin{figure}[htp]
\begin{center}
 \includegraphics[width=0.7\textwidth]{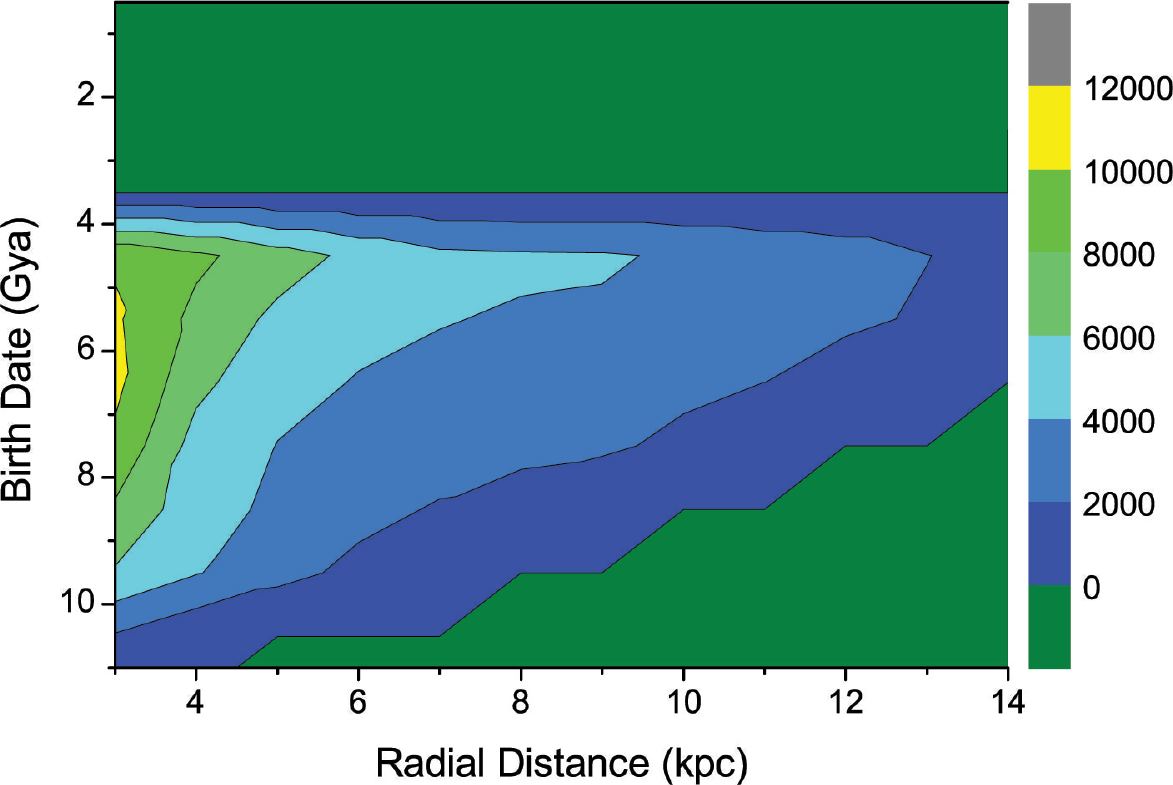}
  \caption{The distribution of birth dates of the number of habitable planets per pc$^{2}$ (color scale) at the present day \citep[Model~4 from][]{2011AsBio..11..855G} as a function of radial distance.  Planets are not considered habitable if they have ages $<4$ Gyr. The inner Galaxy attains habitable planets earlier than at the solar radius ($R_\odot=8$ kpc), and at the present day, the inner Galaxy at $R=2.5$ kpc, has the greatest area density of habitable planets.} 
  \label{Fig10_from_2011AsBio..11..855G}
\end{center}
\end{figure}

\begin{figure}[htp]
\begin{center}
 \includegraphics[width=0.7\textwidth]{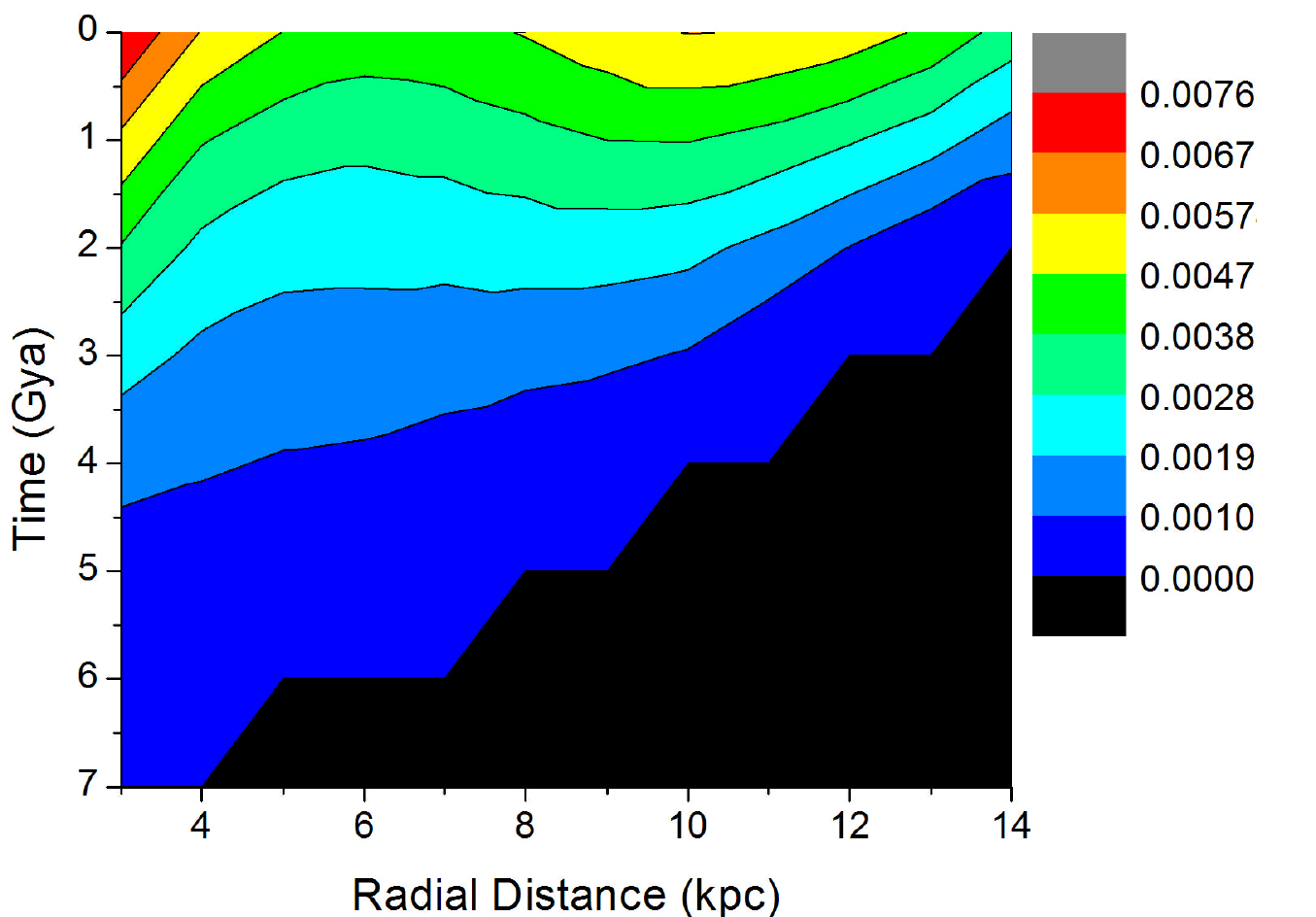}
  \caption{The fraction of stars with a habitable planet (color scale) as a function of time and radial distance over all epochs \citep[Model~4 from][]{2011AsBio..11..855G}.} 
  \label{Fig12_from_2011AsBio..11..855G}
\end{center}
\end{figure}

Figure~\ref{Fig10_from_2011AsBio..11..855G} from \cite{2011AsBio..11..855G} plots the area density (planets per pc$^2$) of the birth dates of habitable planets at the present day.  We found that the inner Galaxy hosts the greatest density of habitable planets, and that the Earth is not in the most habitable region of the Milky Way.  Furthermore, examining the fraction of stars with a habitable planet over all epochs, the region most favorable for life is the inner Galaxy (Figure~\ref{Fig12_from_2011AsBio..11..855G}).  However, the fraction of stars with a habitable planet is only a factor of $\sim2$ higher in the inner Galaxy at recent epochs than the solar radius at 8 kpc (for instance, see the last 1 Gyr in Figure~\ref{Fig12_from_2011AsBio..11..855G}).  Stars form earlier in the inner Galaxy, which in turn build up a large number of planets through increased metallicity abundances before planet formation occurs at higher galactocentric radii (see the metallicity distribution in Figure~\ref{Fig1_from_2011AsBio..11..855G}).  The high planet density and older mean planet age in the inner Galaxy outweighs the negative impact of SNe. High metallicity abundances can hinder terrestrial planet formation through the formation of gas giants.  Over time, the inner Galaxy builds up a large number of terrestrial planets, and likely only hinders the formation of terrestrial planets due to high metallicity abundances at later epochs.  In summary, \citet{2011AsBio..11..855G} find that the inner Galaxy is favorable to both metrics of habitability: 1) the number density, and 2) the fraction of stars that host a habitable planet.

The Monte Carlo modeling exercise of \cite{2011AsBio..11..855G} was extended in \cite{2015AsBio..15..683M} to consider not only habitability for complex land-based life, but also habitability over the extended timescales thought to be required for the emergence of intelligence.  Depending on the epoch time and location within the Galaxy, habitable planets are at greater or lesser risk of experiencing sterilizing SNe events.  Gaps between such events provide windows of opportunity for complex/intelligent life to evolve. \cite{2015AsBio..15..683M} investigates the spatial and temporal distribution of these opportunities to gain an understanding of the way the likelihood of complex/intelligent life arising may have varied over cosmic timescales in different regions of the Galaxy.

\begin{figure}[htp]
\begin{center}
 \includegraphics[width=0.7\textwidth]{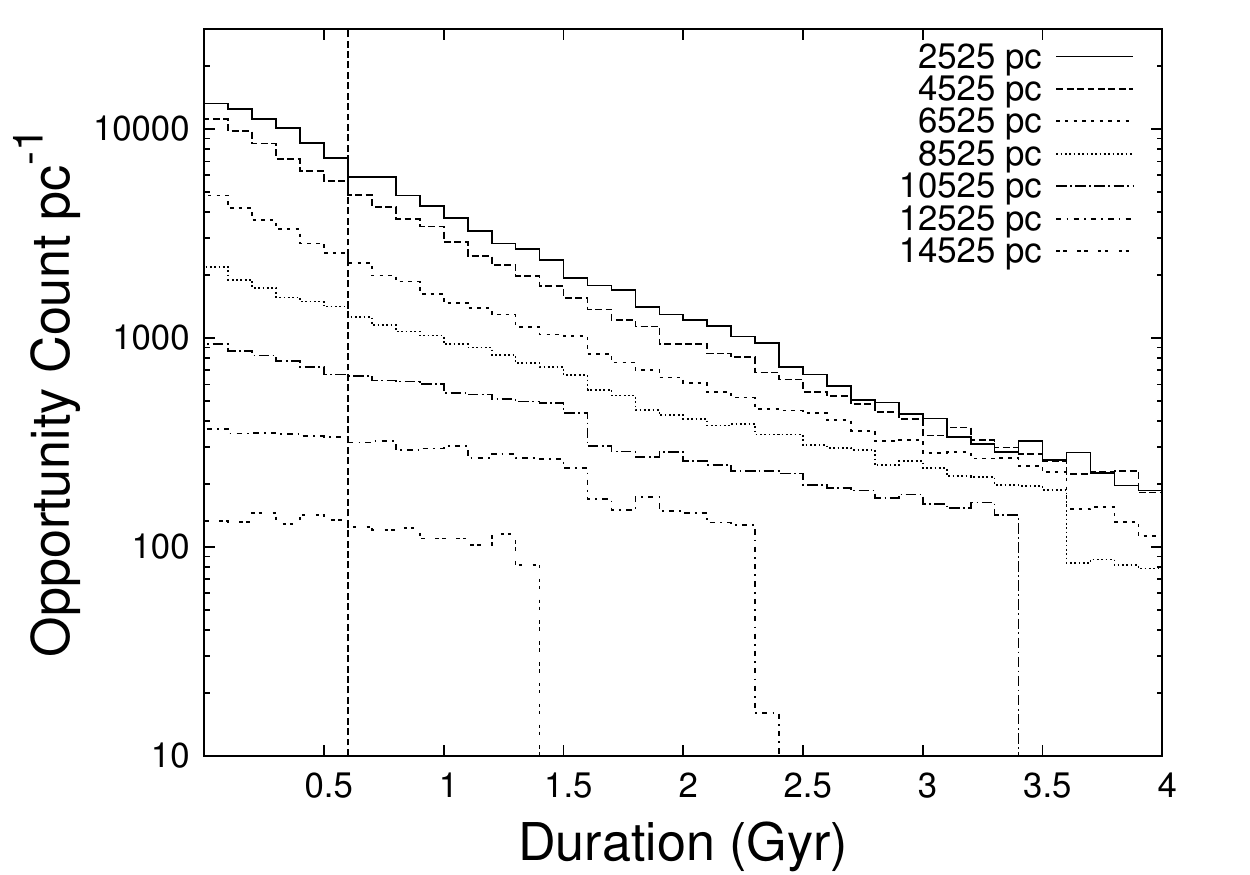}
  \caption{Figure 9 from \citet{2015AsBio..15..683M} that plots histograms of the number of opportunities for the emergence of complex/intelligent life of different durations, summed over all epochs, for a range of different galactocentric radii.  The dashed vertical line crossing the horizontal axis at 0.6 Gyr corresponds to the opportunity duration experienced when intelligence arose on Earth. The opportunity count pc$^{-1}$ is calculated by summing the total number of opportunities of a given duration in a radial bin (an annulus) and then dividing by the bin width (50 pc). This yields the absolute opportunity counts within each annular region.} 
    \label{Fig9_from_2015AsBio..15..683M}
\end{center}
\end{figure}

In terms of the spatial distribution of opportunities, this was found to mirror the findings of \cite{2011AsBio..11..855G}.  Regardless of the duration of undisturbed habitability presumed for an opportunity, the greatest number of opportunities of that duration are always expected in the inner Galaxy.  This can be seen in Figure~\ref{Fig9_from_2015AsBio..15..683M} taken from \cite{2015AsBio..15..683M}, which shows histograms of the number of opportunities of different durations for a range of different distances from the Galactic center.   For all duration lengths, the number of opportunities is at its maximum in the inner Galaxy and decreases monotonically with increasing galactocentric radius.  Whether habitability is defined over relatively short or long timeframes, the significantly higher number density of habitable planets outweighs the impact of more frequent SNe events and the inner Galaxy is always favored.

\begin{figure}[htp]
\begin{center}
 \includegraphics[width=0.7\textwidth]{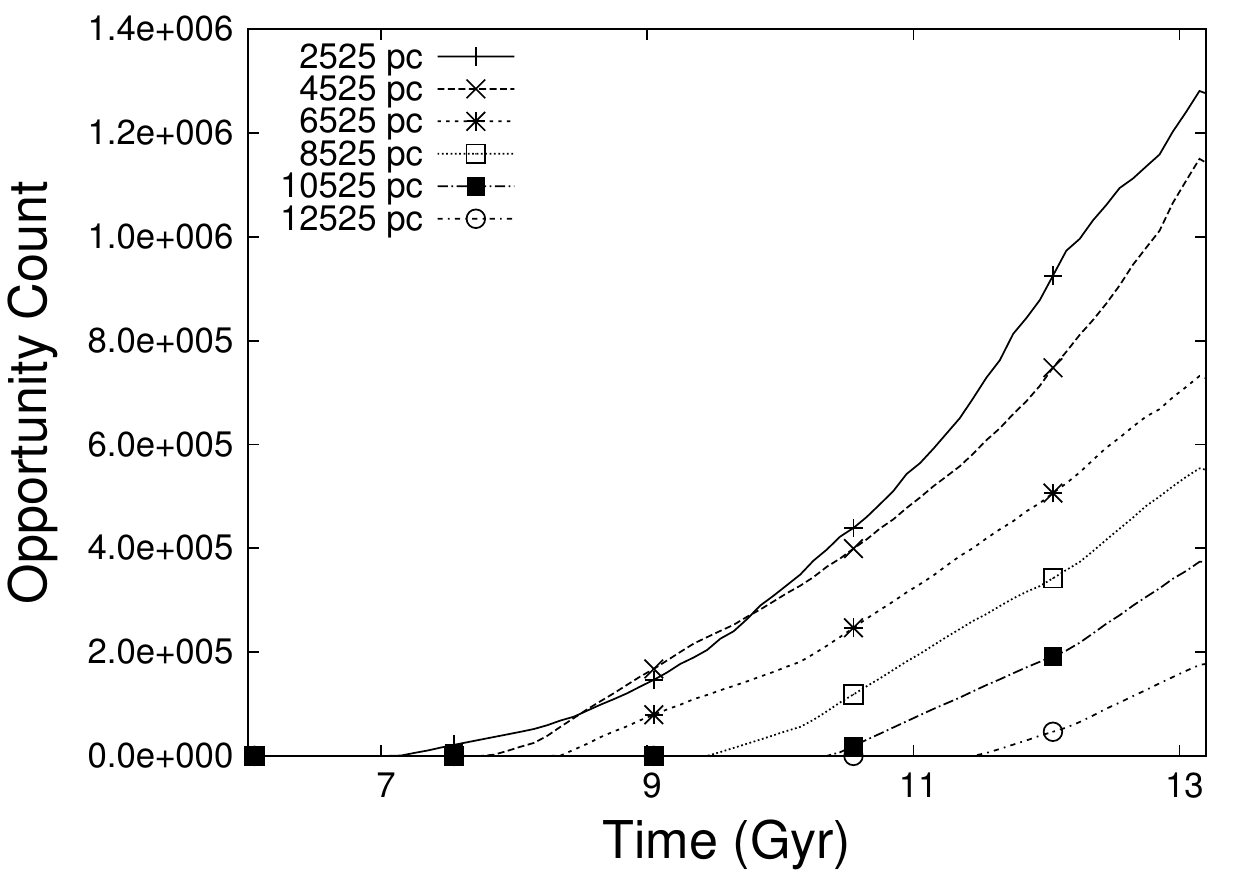}
  \caption{Figure 10 from \citet{2015AsBio..15..683M} that plots  the number of active opportunities for the emergence of complex/intelligent life versus time since the formation of the Galaxy, for a range of different galactocentric radii.} 
    \label{Fig10_from_2015AsBio..15..683M}
\end{center}
\end{figure}

Due to the inside-out star formation history of the Galaxy, the first habitable planets appear in the inner Galaxy and at progressively later epochs with increasing galactocentric radius.  At all radii, the number density of habitable planets climbs steadily over time.  However, since the inner Galaxy has a ``head start'' due to the inside-out formation history, it is found that there are always a greater number of habitable planets toward the inner Galaxy, in all epochs.  Even accounting for the higher rates of sterilizing SNe events in the inner Galaxy, across all epochs there are more windows of opportunity for complex life to emerge in the inner Galaxy than at higher radii.  This can be seen in Figure~\ref{Fig10_from_2015AsBio..15..683M} taken from \cite{2015AsBio..15..683M}, which plots a count of the total number of active opportunities (defined as habitable planets experiencing a time gap between SNe of at least 2.15 Gyr) as a function of time since the formation of the Galaxy, for various galactocentric radii.  At higher radii, opportunities do not begin to occur until later epochs, due to insufficient planetary age.  At all radii the number of opportunities increases steadily over time, reaching its maximum today.  This trend can be expected to continue for several billion years into the future as the metallicity increases throughout the Galactic disk, thus supporting higher planet formation rates.

A key point to note from Figure~\ref{Fig10_from_2015AsBio..15..683M} is that the level of opportunity for the emergence of complex life at the present time at Earth's galactocentric radius was matched in the distant past in regions closer to the Galactic center.  Specifically the modeling is suggestive of there existing a level of habitability in the inner Galaxy around 2 Gyr ago that was similar to that in our local neighborhood today.  Furthermore, the very first opportunities for complex life began to appear in the Galaxy more than 5 Gyr ago, \textit{before the formation of the Earth}.  We discuss these findings further in Section \ref{Section_HUBE}.

The chemical evolution of the Milky Way determines the regions of the Galaxy that host habitable planets over time. \cite{2014MNRAS.440.2588S} model the habitability of both the Milky Way and Andromeda (M31). In their work, the authors compare the effects of radial gas flows on the number of stars that host a habitable planet.  By including radial gas flows, the number of habitable planets increases by 38\% in comparison to modeling the chemical evolution without radial gas flows, and they find that the location with the highest number of stars with planets is at a galactocentric radius of $R=8$ kpc.  This result is similar to that found in \cite{2004Sci...303...59L}, which may be due to the method used to model the dangers of SNe events to planets in the Milky Way.  

All of the GHZ models described above~\citep{2001Icar..152..185G,2004Sci...303...59L,2008SSRv..135..313P,2011AsBio..11..855G,2014MNRAS.440.2588S,2015AsBio..15..683M} have employed major observational properties to model the GHZ; however, they have excluded the effects of stellar kinematics.  \cite{2015arXiv151101786F} advanced a model of the habitability of the Milky Way using high resolution N-Body cosmological simulations to understand the effects of the evolution of the Local Group on the Milky Way.  Thus, the work is able to give an account of the mass assembly history of the Milky Way in comparison to the other models that largely assume azimuthal symmetry\footnote{As an example of azimuthal symmetry, if you were to stand in the center of the Galaxy and studied an observable property, such as the number of stars in a region at a fixed distance, you would obtain the same number of stars if you were to rotate your body and look at the same sized region at the same fixed distance.}. \cite{2015arXiv151101786F} report that the GHZ of the Milky Way is between 2 and 13 kpc and that there are a large number of habitable planets in the inner Galaxy.  The probability of a star hosting a habitable planet increases with galactocentric radius, whereas the number density of habitable planets decreases with radius. Another interesting paper by \cite{2016MNRAS.459.3512V} uses an N-Body simulation of an isolated Milky Way-like galaxy to study dynamical effects on the GHZ. They trace the locations of the stellar mass particles (roughly the mass of a star cluster) over time to determine the timespan within which they reside in a habitable region, as defined by a stellar number density and star formation rate below a particular threshold. \cite{2016MNRAS.459.3512V} find that the most habitable region of the Milky Way at the present day is towards the outskirts, and less than 1\% of habitable planets are found near the solar radius at 8--10 kpc \citep[see also][]{Vukotic2017}.  This result suggests that the Earth's location in the Galaxy is atypical, and it is inconsistent with the 7--9 kpc GHZ found in \cite{2004Sci...303...59L} or \cite{2014MNRAS.440.2588S}. As  noted in \cite{2016MNRAS.459.3512V}, this is likely due to the thresholded stellar density and star formation rate assumed to be required for a habitable environment, and in addition, a possible cause might be the dynamical effects that cause radial mixing of stellar populations with more metal rich systems migrating outwards from the inner regions of the Galactic disk.

\begin{table}[!htp]
\centering
\caption{Comparison of selected literature sources. Major observables may include: the star formation history, metallicity evolution, stellar number density distribution, and other properties. Lethal events refer to those dangers to habitable planets in the models. Assessment of most habitable region at the present day ($z\sim0$) refers to quoted values in the studies.}
\begin{small}
\begin{tabular}{|c|c|c|c|} \hline
Source&Model Type&Lethal Events&Assessment at $z\sim0$\\ \hline
\hline
\citet{2001Icar..152..185G}&\makecell{Major \\Observables}&N/A&\makecell{Thin disk near the \\Sun for formation of \\Earth-like planets}\\ \hline
\citet{2004Sci...303...59L}&\textquotedbl&SNe \& Hot Jupiters&$R=7-9$ kpc\\ \hline
\citet{2008SSRv..135..313P}&\textquotedbl&\textquotedbl&Entire disk\\ \hline
\citet{2011AsBio..11..855G}&\textquotedbl&SNII, SNIa \& Hot Jupiters&\makecell{Highest density at \\$R\approx 2.5$ kpc}\\ \hline
\citet{2014MNRAS.440.2588S}&\textquotedbl&\textquotedbl&$R=8$ kpc\\ \hline
\citet{2015AsBio..15..683M}&\textquotedbl&\textquotedbl&$R\approx 2.5$ kpc\\ \hline
\citet{2015arXiv151101786F}&N-Body&\textquotedbl&$R=2-13$ kpc\\ \hline
\citet{2016MNRAS.459.3512V}&\textquotedbl&\makecell{SNe approximated by \\ stellar number density \\and star formation rate thresholds}&\makecell{Predominantly the \\outskirts, $R\sim16$ kpc}\\ \hline 
\end{tabular}
\end{small}
\label{tab:comparison}
\end{table}

Table~\ref{tab:comparison} compares the GHZ literature discussed above, which is not a complete summary of all of the papers on the habitability of the Milky Way.  In the table, we show selected literature, the type of model, what lethal events to habitability were considered, and the region that is predicted to be the most habitable at the present day.  Most of the studies list numerous caveats and have differing metrics of habitability (e.g., the location of the greatest number density of habitable planets vs. the greatest fraction of stars with habitable planets). For instance, \cite{2004Sci...303...59L} find the location with the greatest probability of a star hosting a habitable planet to be between $R=7-9$ kpc, \citet{2008SSRv..135..313P} finds a maximum probability at $R=10$ kpc, and \citet{2011AsBio..11..855G} find this probability to be highest in the inner Galaxy (but only by a factor of $\sim2$ higher than at $R=8$ kpc). However, \citet{2008SSRv..135..313P} and \citet{2011AsBio..11..855G} find that the number density of habitable planets is greatest in the inner Galaxy, and \cite{2004Sci...303...59L} does not calculate this number density.  Given the different metrics and methods, we refer the reader to the papers in Table~\ref{tab:comparison}, and note that we have endeavored to interpret the results of these studies as accurately as possible. 

From Table~\ref{tab:comparison} there are numerous predictions regarding the most habitable region of the Milky Way.  Excluding \citet{2001Icar..152..185G} (because dangers to planets were not modeled) and \citet{2015AsBio..15..683M}  \citep[because it is an extension of][and has a similar methodology]{2011AsBio..11..855G}, three papers suggest that nearly the entire disk is habitable, two papers suggest that the region encompassing the solar radius ($R_\odot=8$ kpc) is the most habitable, and one paper suggests that the outer disk is the most habitable region at the present day.  Thus, there is a divergence in the literature, which we believe to be due to the fact that model outcomes are heavily reliant on model assumptions and methodologies \citep[as was pointed out in][]{2008SSRv..135..313P}.  In what follows, we describe some of the reasons for the divergence in the literature.

\cite{2001Icar..151..307L} informed the probability of forming an Earth-like planet in \cite{2004Sci...303...59L}, where Earth-like planet formation was strongly correlated with metallicity (but declines when high metallicity environments inhibit Earth-like planet formation, due to Hot Jupiters).  From simulations of planet formation, \cite{2008SSRv..135..313P} argued for a constant probability of forming Earth-like planets as a function of metallicity. \citet{2011AsBio..11..855G} combined the planet-metallicity correlation for Hot Jupiters~\citep{2005ApJ...622.1102F} with statistics from the formation of Earth-mass planets in models of planet formation. \citet{2014MNRAS.440.2588S} and \cite{2015arXiv151101786F} use similar assumptions as \citet{2008SSRv..135..313P}.  \citet{2016MNRAS.459.3512V} assign star particles a metallicity abundance from a Gaussian distribution and then determine if the particle is habitable based on whether the value is above that of a sample of detected Earth-like exoplanet host stars. With the exception of \cite{2004Sci...303...59L}, the studies generally assume that small planet formation is weakly dependent on metallicity (as shown in Figure~\ref{Fig3_from_2012Natur.486..375B}).   Although these methods slightly differ, over recent epochs, the models produce a large number of planets throughout the Galactic disk that can be examined to determine whether they are habitable. Thus, it is unlikely that the relationship between metallicity and terrestrial planet formation is responsible for the divergence in the model predictions.

Another difference in the literature is the metric that is used to determine whether a planet is habitable. Most of the papers define the GHZ as a probability function that takes as input the radial distance, and time.  With this probabilistic formulation, it is straightforward to assume that the danger to planets due to proximity to SNe can be modeled as a function of the SN rate in the region.   In particular, all of ~\citet{2004Sci...303...59L, 2008SSRv..135..313P, 2014MNRAS.440.2588S,2015arXiv151101786F, 2016MNRAS.459.3512V} assume that the danger due to SNe is normalized to a particular SNe rate (typically that of the solar neighborhood).  For instance, if we let $RSN$ be the SN rate integrated over the past 4 Gyr at the solar neighborhood, then \cite{2004Sci...303...59L} assigns a probability value of surviving a SN, $P_{SN}$=1 when the rate is $\leq0.5RSN$, and $P_{SN}=0$ when the rate is $\geq 4RSN$. As mentioned in \cite{2004Sci...303...59L}, this normalization is somewhat arbitrary, as our knowledge of the effects of transient radiation events on the atmosphere and biosphere are not well established. This formulation is conducive to producing hard inner boundaries on the GHZ, as regions will not permit any habitable planets when the SN rate is more than $4\times$ the local neighborhood rate.  In contrast to these methods, the dangers due to SNe in \cite{2011AsBio..11..855G} were calculated by examining the times at which individual stars were sufficiently nearby a SN event, as determined by a distribution of sterilization distances for SNII and SNIa, in combination with using the Earth's history as a template for when SN events will negatively influence the biosphere.  This may more carefully quantify the dangers to planets as a function of SN event rate, and more easily allows for studying habitability above and below the midplane, and not only as a function of galactocentric radius (and time).  Additionally, it is less conducive to producing hard boundaries on the habitable regions of the Galaxy, although it still suffers from our incomplete understanding of the effects of transient radiation events on habitability.  But it is a key factor in explaining why \cite{2011AsBio..11..855G} find that the inner Galaxy is the most habitable region at the present day; even though the habitability of many planets do not survive SN explosions, a large number of planets still withstand these lethal events.  A study that compares the differences between SN lethality modeling methods would help illuminate why the models give differing predictions.

\section{The Habitability of Other Galaxies}

The habitability of the Milky Way has understandably received greater attention than other galaxies, partly because it is our home, but predominantly because we have much more observational data at our disposal with which to calibrate our models of habitability.  However, consideration of other galaxies is essential to characterizing the habitability of the wider Universe.  Furthermore, observational data from distant galaxies may provide a valuable window into the past, and thus help to improve our understanding  of the history of the Milky Way.   The habitability of other galaxies is an emerging area of research that has generated a number of recent publications including  \cite{suthar2012}, \cite{2013RMxAA..49..253C}, \cite{2014MNRAS.440.2588S}, \cite{2015arXiv151101786F} and \cite{2015ApJ...810L...2D}. 

\cite{2013RMxAA..49..253C} and \cite{2014MNRAS.440.2588S} consider galaxy M31 (Andromeda); the nearest spiral galaxy to the Milky Way.  The presumed similarities in structure  and formation history of all spiral galaxies allows similar methodologies to be applied as discussed previously for the Milky Way, but modified to take account of M31's different mass; nearly twice that of the Milky Way.  M31 makes an ideal test-bed for understanding the habitability of even larger spiral galaxies.  \cite{2013RMxAA..49..253C}  apply a similar methodology as \cite{2004Sci...303...59L} and estimate M31's GHZ. They find that when considering the highest surface density of habitable planets, the GHZ lies between 3--7 kpc. The region with the greatest number of habitable planets, however, is found between 12--14 kpc.

 \cite{2014MNRAS.440.2588S} extend the work of \cite{2013RMxAA..49..253C} to include radial gas flows.  They apply their model to both M31 and the Milky Way, predicting similar radii of maximum habitable planet density as  \cite{2013RMxAA..49..253C}  but slightly higher total counts for habitable planets.  The fact that they applied precisely the same methodology to both M31 and the Milky Way allows for a direct comparison of the two galaxies.  They estimate that the maximum habitable planet density occurs at 8 kpc radius for the Milky Way and 16 kpc for the more massive M31. 
 
One may wonder about the near-future habitability of M31 and the Milky Way. The potential merger of M31 and the Milky Way in $\sim$ 4 Gyr may result in the formation of a single elliptical galaxy with a very different habitability distribution and trajectory than either of its constituent progenitor spiral galaxies.  For this reason we cannot extrapolate the habitability of the Milky Way or M31 too far into the future.

The habitability of elliptical galaxies has been considered by \cite{suthar2012}.  This work is more speculative because it is necessary to make assumptions about star formation rates, metallicity distributions and Earth-like exoplanet formation that are less constrained by observational data than our Galaxy or M31.  Based primarily on statistical considerations of metallicity distributions, \cite{suthar2012} conclude that the elliptical galaxies M87 and M32 may host a multitude of habitable Earth-like planets. However, this is based on numerous assumptions, so a more reasonable conclusion may be that certain regions of these elliptical galaxies can be expected to contain stars following metallicity distributions that are not completely disjoint from our region of the Milky Way, so we may reliably infer that small rocky planets will be present - but in what fraction of planetary systems is uncertain.  Nevertheless, the fact that we expect the existence of Earth-like planets in elliptical galaxies is important because such galaxies are abundant in the Universe and are believed to contain generally older stars.  Therefore they potentially host a great number of habitable planets on which there have been long durations undisturbed by sterilization events, offering many opportunities for the emergence of complex life.

Consideration of other galaxies and galaxy types has been discussed in the work of \cite{2015arXiv151101786F} and \cite{2015ApJ...810L...2D}.  In \cite{2015arXiv151101786F}, high resolution cosmological simulations of galaxy formation in the Local Group are used to map out the spatial and temporal behavior of the Milky Way and M33 GHZs, in addition to tidal streams and satellite galaxies, allowing for more generalized three-dimensional structures that may not possess azimuthal symmetry.  They conclude that each galaxy's GHZ depends critically on its evolutionary history, in particular on its accretion history.  \cite{2015ApJ...810L...2D} propose a framework that allows the consideration of all galaxy types, with a view to understanding which types offer the greatest degree of habitability.  Studying $\sim 100,000$ galaxies from the Sloan Digital Sky Survey~\citep{2011AJ....142...72E}, their work suggests that giant metal-rich elliptical galaxies at least twice as massive as the Milky Way could potentially host many thousands of times more habitable planets than the Milky Way.  Due to the large sample of galaxies studied, they primarily rely on fewer observable properties than used in studies of the Milky Way to derive their estimates of habitability.  Their approach makes a number of simplifying astrophysical assumptions and approximations, such as a homogeneous distribution in galaxies of stars/SNe and that the volume scales with the total stellar mass, together with simplistic assumptions about the effects of sterilizing radiation events on the evolution of complex life. Nonetheless, their work is a good step towards understanding habitability from a cosmological context.  Future refinements to simulation methodologies and astrophysical assumptions in star/galaxy formation processes can be expected to yield steady improvements to our understanding of GHZ evolution in all galaxy types and hence the Universe at large.

\section{Transient Radiation Events}
Research on the GHZ has focused on SNe events.  However, there are other transient radiation events that may inhibit life on planets in the Milky Way. In particular, Gamma-Ray Bursts (GRBs) have a beamed emission that can affect planets on the order of 1 kpc, potentially causing mass extinction events to many planets in the Milky Way~\citep{1995ApJ...444L..53T,2005ApJ...634..509T,2005ApJ...622L.153T,2011AsBio..11..343M,2014PhRvL.113w1102P,2015ApJ...810...41L,2016GowanlockGRBApJ}. For overviews on the origin of GRBs, see~\citet{2002ARA&A..40..137M}, \citet{2006ARA&A..44..507W}, and \citet{2009ARA&A..47..567G}, and references in the above mentioned articles.  Furthermore,   active galactic nuclei (AGN) may be dangerous to planets in the inner Galaxy and bulge~\citep{1981CLARKE}.  Both AGN and GRBs have not been modeled with SNe in studies of the GHZ; rather, these dangers have often been modeled separately, for instance to study the dangers of GRBs to the Earth~\citep{2004IJA:240775,2005ApJ...622L.153T,2005ApJ...634..509T,Melott2009,2015AsBio..15..207T}. AGN are likely not very dangerous to planets in the Milky Way, as the range in which their radiation would impact the habitability of planets is limited to the innermost regions of the Galaxy. Here, we briefly summarize recent findings on the effect of GRBs to the habitability of planets in the Milky Way.

\cite{2014PhRvL.113w1102P} modeled the effects of GRBs to planets within the disk of the Milky Way. They find that long GRBs are the most dangerous GRB type, and that short GRBs are a fairly negligible source of transient radiation.  Long GRBs have a metallicity dependence, as they are predominantly found in low metallicity galaxies~\citep{2006Natur.441..463F}, and thus are typically likely to occur at high redshift.  Using the metallicity-dependent GRB rate and moderate GRB fluence lethality threshold values, \cite{2014PhRvL.113w1102P} find that over the past 1 Gyr leading up to the present day, there is a 60\% chance of a planet at the Earth's galactocentric radius to be irradiated by a long GRB. \cite{2015ApJ...810...41L} find that there is roughly one long GRB at Earth's radius every 500 Myr, and \cite{2016GowanlockGRBApJ} reports that $\sim35\%$ of planets at the solar radius are within the  beam of a long GRB over the past 1 Gyr. 

\begin{figure}[htp]
\begin{center}
 \includegraphics[width=0.7\textwidth]{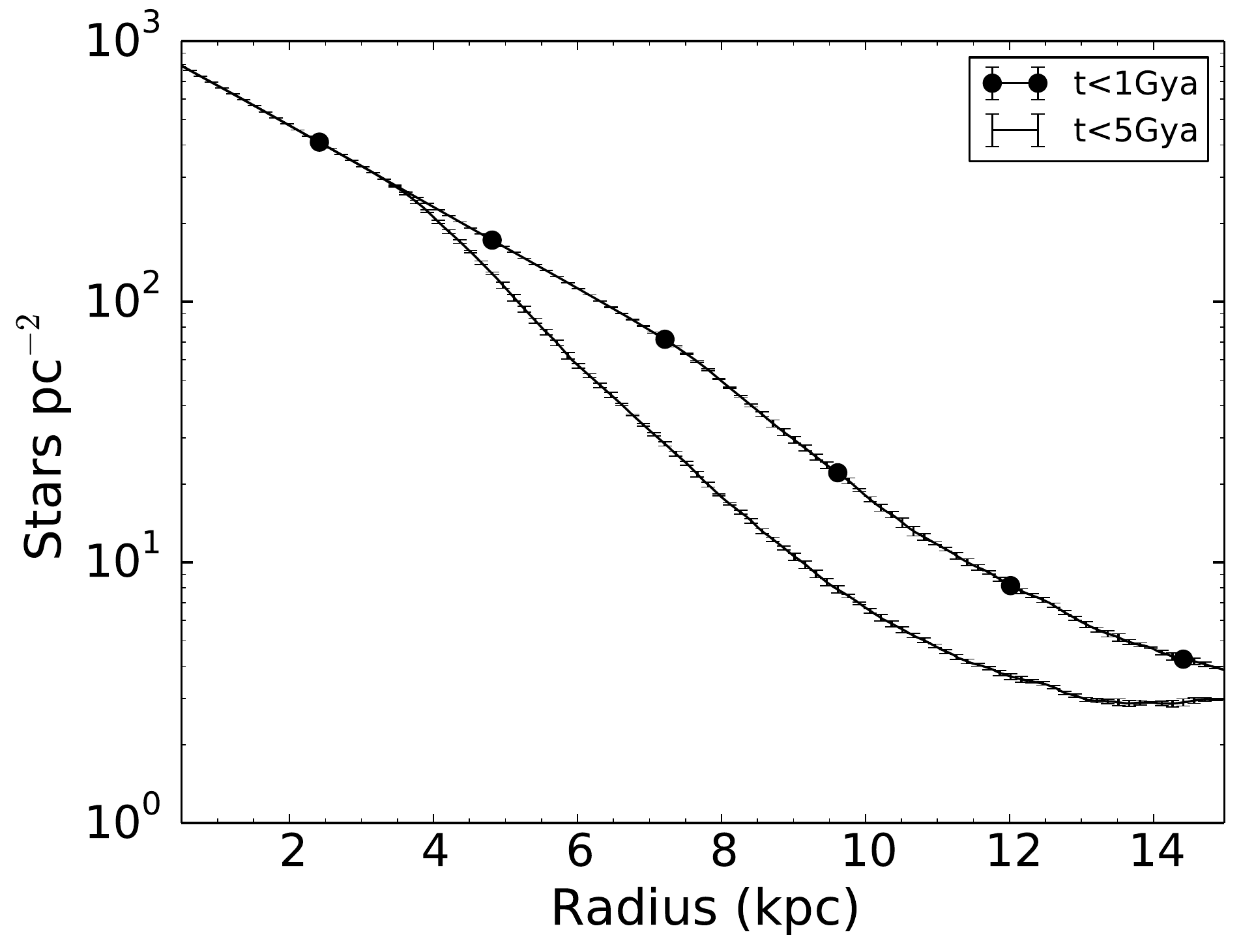}
  \caption{The surface density of stars that are not within the beam of a GRB over the past 1 Gyr and 5 Gyr, as a function of galactocentric radius \citep[Model~2 from][]{2016GowanlockGRBApJ}. Reproduced by permission of the author.} 
  \label{Fig8_from_2016GowanlockGRBApJ}
\end{center}
\end{figure}

Without comparing the individual model assumptions of these works we simply note that the chemical evolution of the Milky Way may be sufficiently advanced to quench long GRB formation, as these GRBs are found in low-metallicity environments~\citep{2006Natur.441..463F}, and have primarily been found in galaxies less massive than the Milky Way~\citep{2013ApJ...773..126J}. Thus, it may be the case that long GRBs may not strongly contribute to the sources of transient radiation hazards in the Galaxy.  Furthermore, assuming a linear dependence between metallicity and GRB formation (low metallicity produces more GRBs), \cite{2016GowanlockGRBApJ} reports that the only environment that would produce GRBs at the present day is the galactic outskirts. However, at higher redshift, there are likely to be more long GRBs at lower galactocentric radii that may reduce the habitability of planets at that time. Figure~\ref{Fig8_from_2016GowanlockGRBApJ} plots the area density of stars that survive a long GRB over the past 1 Gyr and 5 Gyr from Model~2 in \cite{2016GowanlockGRBApJ}.  From the plot, the inner Galaxy has the highest number density of stars that survive a long GRB event, in part because the chemical evolution has quenched GRB formation in the region.  The overlap in the 1 Gyr and 5 Gyr curves at $R\lesssim 4$ kpc, indicates that no GRB formation has occurred in that region over the past 5 Gyr. GRB events should be incorporated into calculations of the habitability of the Milky Way; however, SNe may still be the dominant source of transient sterilizing radiation in the Galaxy.

\section{The Habitability of the Galaxy Before the Earth} \label{Section_HUBE}
The majority of the studies of habitability on galactic scales focus on the present day. This is likely due to a few reasons. First, astronomy and astrophysics have intensely studied the structure and properties of the Milky Way. Therefore, the morphology and characteristics of the Milky Way at the present day are well known.  In contrast, at earlier epochs, the overall characteristics of the Milky Way are less well constrained, and we rely on models that reproduce the major observable properties we see at present.  Additionally, as our civilization exists today, notions of habitability are most relevant at the present day (such as estimating the distribution of the birth dates of planets that are habitable now in Figure~\ref{Fig10_from_2011AsBio..11..855G}).  Lastly, due to the many uncertainties in assessing habitability described heretofore, it is difficult to validate the habitability of the Galaxy in recent epochs, and even more challenging (or potentially impossible) at earlier epochs.  In what follows, we appraise the habitability of the Galaxy before the Earth formed in the Milky Way 4.5 billion years ago.        

From Figure~\ref{Fig1_from_2011AsBio..11..855G} (upper panel), it is clear that the inner Galaxy had a burst of star formation that has declined with time. Furthermore, the stars initially form in the inner Galaxy and the stellar mass builds outwards with time.  Since the Milky Way would have been a more compact object and thus more dense on average at earlier epochs, SNII that explode shortly after they form, and prompt SNIa, would have had a catastrophic impact on the habitability of the inner Galaxy.  However, with the decline in the star formation rate over time, the planets that formed in the inner Galaxy are eventually able to sustain complex life \citep[according to][]{2011AsBio..11..855G,2015AsBio..15..683M}. If we examine earlier epochs in  Figure~\ref{Fig10_from_2011AsBio..11..855G}, there are many planets in the inner Galaxy at $R\sim2.5$ kpc that formed 9 Gyr ago that are habitable today. That is roughly double the lifetime of the Earth. Therefore, given the age distribution of stars in the Milky Way, it is likely that the first habitable planets began to appear $\gtrsim 10$ Gyr ago. In fact, Kapteyn's star is thought to host two super-Earth planets where the star is estimated to be between $10-12$ Gyr old~\citep{2014MNRAS.443L..89A}.  Therefore, if we only consider the lethality of SNe to biospheres, then the Galaxy is likely to have hosted many habitable environments before the Earth formed.

This conclusion is supported by the findings of \cite{2015AsBio..15..683M}, as already discussed in Section \ref{Section_Hab_MW}.   We saw in Figure~\ref{Fig10_from_2015AsBio..15..683M} that the number of opportunities for the emergence of complex life in the Milky Way has been steadily increasing over time, at all galactocentric radii.  We might broadly equate the likelihood of complex life emerging at any given epoch with the number of active opportunities in that epoch.  Given that complex life is known to have emerged at least once near the present time at Earth's galactocentric radius, it is worthwhile considering in what epoch a similar number of opportunities were available in other regions of the Galaxy.  Figure~\ref{Fig10_from_2015AsBio..15..683M} suggests a similar likelihood for the emergence of complex life existed in the inner Galaxy around 2 Gyr ago to that seen at Earth's galactocentric radius today.  We should not infer from this that complex life \textit{did} in fact emerge previously within the inner Galaxy, since Earth might represent a statistical outlier event.  But if it did, we might expect such life to be considerably older than ourselves.  Indeed, it may pre-date the very formation of the Earth.  As explained above, with the first habitable planets appearing more than 10 Gyr ago, there was a subsequent duration of more than 5.5 Gyr on those planets before the time that the Earth formed.  This is a substantially longer duration than the $\sim$ 4 Gyr that it took on Earth for complex life to appear after the planet's formation. 

If we include the dangers of long GRBs to planets before the Earth formed (Figure~\ref{Fig8_from_2016GowanlockGRBApJ}), then 5 Gyr ago, there were numerous stars that survive GRB events in the inner Galaxy. And even if we consider Model~1 in \cite{2016GowanlockGRBApJ} that favors hosting GRBs primarily in the inner Galaxy over all epochs, the number density of stars that survive a GRB event over the past 5 Gyr is roughly the same between $2.5<R<8$ kpc.  Assuming that a planet is not deemed uninhabitable if it is within the beam of a GRB once every 5 Gyr, then the entire disk of the Galaxy would have hosted a large number of habitable planets before the Earth formed. This is consistent with \cite{2015ApJ...810...41L} that report that even though the GRB rate is higher at $z>0.5$ (roughly 5 Gyr ago), many planets are expected to survive long GRBs in the Milky Way and in other galaxies examined in samples from the Sloan Digital Sky Survey.  If we combine the dangers of long GRBs and SNe, the Galaxy is likely to have hosted habitable planets throughout the Galactic disk, well before the Earth formed, with the exception of the very early Galaxy that would have hosted too many SNe, and likely long GRBs that would not have been quenched by sufficiently high metallicity abundances at that time.

\section{Conclusions and Future Outlook}
The habitability of the Milky Way is still an open question.   The study of habitability on large spatial scales necessitates that we define a habitable environment as one that would be endangered by a transient radiation event that would deplete protective ozone in a planetary atmosphere. With our limited knowledge of the consequences of such an event, we must presume that this would likely cause a mass extinction to surface dwelling complex life. From the exoplanet era, we have learned that a substantial fraction of stars host Earth-sized planets.  These planets do not require a large abundance of metals to form, suggesting that terrestrial planets are located throughout the entire Galactic disk and are of a wide range of ages.  The Milky Way may indeed host a large number of planets that have not been sterilized by transient radiation events, such as SNe and long GRBs; however, predictions vary in the literature and as we are in the early period of the development of this field, it is unlikely that consensus will be reached in the near term.  Future observations that derive a large sample of the compositions of planetary atmospheres and possible signs of biosignatures will help determine a sample of truly habitable planets that can be used to better predict the distribution of habitable environments throughout the Galaxy.  Galactic archaeology studies, starting with reconstructing the history of the solar neighborhood, may lead to a better understanding of the frequency that planets will survive lethal astrophysical events~\citep{2016ApJ...826L...3T, 2016breitschwerdt,2016wallner} and emerging studies that incorporate galactic dynamics will illustrate that the galactic environments in which habitable planets are found may vary widely with time~\citep{2008ApJ...684L..79R}. These new and future discoveries will further complicate our ideas regarding habitability on large spatiotemporal scales, but will surely help us better understand the prospects for life in our Galaxy and the Universe as a whole.

% ==== BIBLIOGRAPHY ============================================================
% ---- use AMS reference format (use file ametsoc.bst included in ...
%      ... http://www.ametsoc.org/pubs/journals/AMS_Latex_V3.0.tar.gz):s
\bibliographystyle{ametsoc}   % ametsoc.bst in local directory
% ---- use BibTeX database file:
% \bibliography{bibliography}

\end{document}